\documentclass[]{spie}  

\usepackage{amsmath,amsfonts,amssymb}
\usepackage{graphicx}
\usepackage[colorlinks=true, allcolors=blue]{hyperref}

\title{XuYi Interferometer Observation Simulator}

\author[a,b,c]{M. Hadjara}
\author[a]{T. Xu}
\author[a]{Y. Sun}
\author[a]{W. Wei}
\author[a]{C. Liu}
\author[a]{H. Zhang}
\author[a]{J. Huang}
\author[a]{C. Zeng}
\author[a]{J. Yuan} 
\author[a]{J. Perdigon}
\author[d,e]{A. Boskri}
\author[f]{S. Lagarde}
\author[f]{R. Petrov}
\author[f]{R. Ligi}
\author[f]{B. Lopez}
\author[f]{F. Millour}
\author[f]{J. Dejonghe}
\author[g]{F. Baron}
\author[h]{J. Wang}
\author[i]{X. Wu}
\author[i]{H. Zhao}
\author[a]{Y. Hou}
\author[a]{Z. Hu}
\author[a]{X. Cui}

\affil[a]{Nanjing Institute of Astronomical Optics \& Technology, National Astronomical Observatories, Chinese Academy of Sciences, Nanjing 210042, China}
\affil[b]{Chinese High Angular Resolution Southern Astronomical Laboratory (CHARSAL), Dept. of Astronomy,  Camino El Observatorio \#1515, Las Condes, Santiago, Chile. Postal Code: 7591245. Casilla 36-D, Santiago - Chile}
\affil[c]{Astro-Photonics Laboratory/Space and Planetary, Exploration Laboratory (SPEL), Dept. of Electrical Engineering, FCFM, Universidad de Chile, Av. Tupper 2007, Santiago, Chile}
\affil[d]{Shanghai Astronomical Observatory, Chinese Academy of Sciences, Shanghai, China}
\affil[e]{LPHEA Laboratory, Oukaimeden Observatory, Cadi Ayyad University/FSSM, BP 239, Marrakesh, Morocco}
\affil[f]{Université de la Côte d’Azur, Observatoire de la Côte d’Azur, CNRS, Laboratoire Lagrange, Bd de l’Observatoire, CS 34229, 06304 Nice cedex 4, France}
\affil[g]{The CHARA Array of Georgia State University, Mount Wilson Observatory, Mount Wilson, CA 91023, USA}
\affil[h]{Institute of High Energy Physics, Chinese Academy of Sciences, Beijing 100049, People's Republic of China}
\affil[i]{Purple Mountain Observatory, Chinese Academy of Sciences, Nanjing 210023, People's Republic of China}

\authorinfo{Further author information: (Send correspondence to M.H.)\\M.H.: E-mail: mhadjara@uchile.cl\\  T.X.: E-mail: txu@niaot.ac.cn}
\begin{document} 
\maketitle

\begin{abstract}
The XuYi Interferometer is the first Chinese long-baseline optical interferometer currently under development by the National Institute of Astronomical Optics and Technology (NIAOT) of the Chinese Academy of Sciences. In preparation for future scientific operations, we developed the XuYi Interferometer Observation Simulator (XIOS), a lightweight MATLAB-based software package designed to assist observers in planning interferometric observations.
The simulator combines target observability calculations, interferometric baseline geometry, delay-line constraints, (u,v)-coverage prediction, and visibility estimation into a single graphical user interface. It incorporates astronomical routines from the MAAT toolbox and adopts visualization concepts inspired by ASPRO2. The software allows users to evaluate observing opportunities, predict interferometric performances, and generate realistic observing configurations before telescope operations.
We present the architecture of the simulator, its principal modules, and representative simulation results obtained for the current three-telescope XuYi interferometric configuration operating in the near-infrared H band. 
\end{abstract}


\section{INTRODUCTION}
\label{sec:intro}  

Optical long-baseline interferometry has become a major observational technique for achieving milli-arcsecond angular resolution beyond the diffraction limit of individual telescopes. Facilities such as CHARA, VLTI, NPOI, and SUSI have demonstrated the scientific potential of combining light from multiple apertures to investigate stellar photospheres, circumstellar disks, interacting binaries, and active galactic nuclei.

China is currently developing its first optical long-baseline interferometer at the XuYi Observatory site. The facility consists of three telescopes connected through vacuum delay lines and beam-combination infrastructures designed for near-infrared observations.

As with all interferometric facilities, efficient observation preparation is essential. Observers must simultaneously evaluate:
\begin{itemize}
\item[•] Target observability from the site,
\item[•] Delay-line limitations,
\item[•] Baseline geometries,
\item[•] Earth-rotation synthesis, 
\item[•] (u,v)-plane coverage,
\item[•] Baseline geometries,Expected interferometric visibilities. 
\end{itemize}
To address these requirements, we developed the XuYi Interferometer Observation Simulator (XIOS), a MATLAB graphical software environment that integrates all these calculations into a unified framework.

\section{THE XUYI INTERFEROMETER}
\label{sec:2}

The XuYi Interferometer is located at the XuYi Observatory operated by NIAOT \cite{xu2024xuYi}.

The current configuration includes three telescopes:
\begin{itemize}
\item[•] T1
\item[•] T2
\item[•] T3
\end{itemize}
forming three independent baselines.

\begin{table}[ht]
\caption{summarizes the principal interferometric parameters currently implemented in the simulator.} 
\label{Tab:1}
\begin{center}       
\begin{tabular}{|l|l|} 
\hline
\rule[-1ex]{0pt}{3.5ex}  \textbf{Parameter} & \textbf{Value} \\
\hline
\rule[-1ex]{0pt}{3.5ex}  Number of telescopes & 3 \\
\hline
\rule[-1ex]{0pt}{3.5ex}  Telescope diameter & 600 mm \\
\hline
\rule[-1ex]{0pt}{3.5ex}  Spectral band & H band \\
\hline
\rule[-1ex]{0pt}{3.5ex}  T1–T2 baseline text & ~61 m \\
\hline
\rule[-1ex]{0pt}{3.5ex}  T1–T3 baseline & ~56 m \\
\hline
\rule[-1ex]{0pt}{3.5ex}  T2–T3 baseline & ~100 m \\
\hline
\rule[-1ex]{0pt}{3.5ex}  Maximum delay-line stroke & 44 m \\
\hline
\end{tabular}
\end{center}
\end{table} 
The corresponding angular resolutions ($\theta$) for the considered observing wavelengths ($\lambda$) and the different baseline lengths ($\mathcal{B}$) are estimated using the well-known interferometric relation:
\begin{equation}
\label{eq:1}
\theta \simeq \frac{\lambda}{\mathcal{B}}
\end{equation}
leading to angular resolutions between approximately $3.4$ and $6.1\,mas$ in H band depending on the baseline considered.

The first dispersed-fringe dataset was successfully obtained in April 2026, demonstrating the functionality of the beam transport and fringe detection systems.

\begin{figure} [ht]
\begin{center}
\begin{tabular}{c} 
\includegraphics[height=10cm]{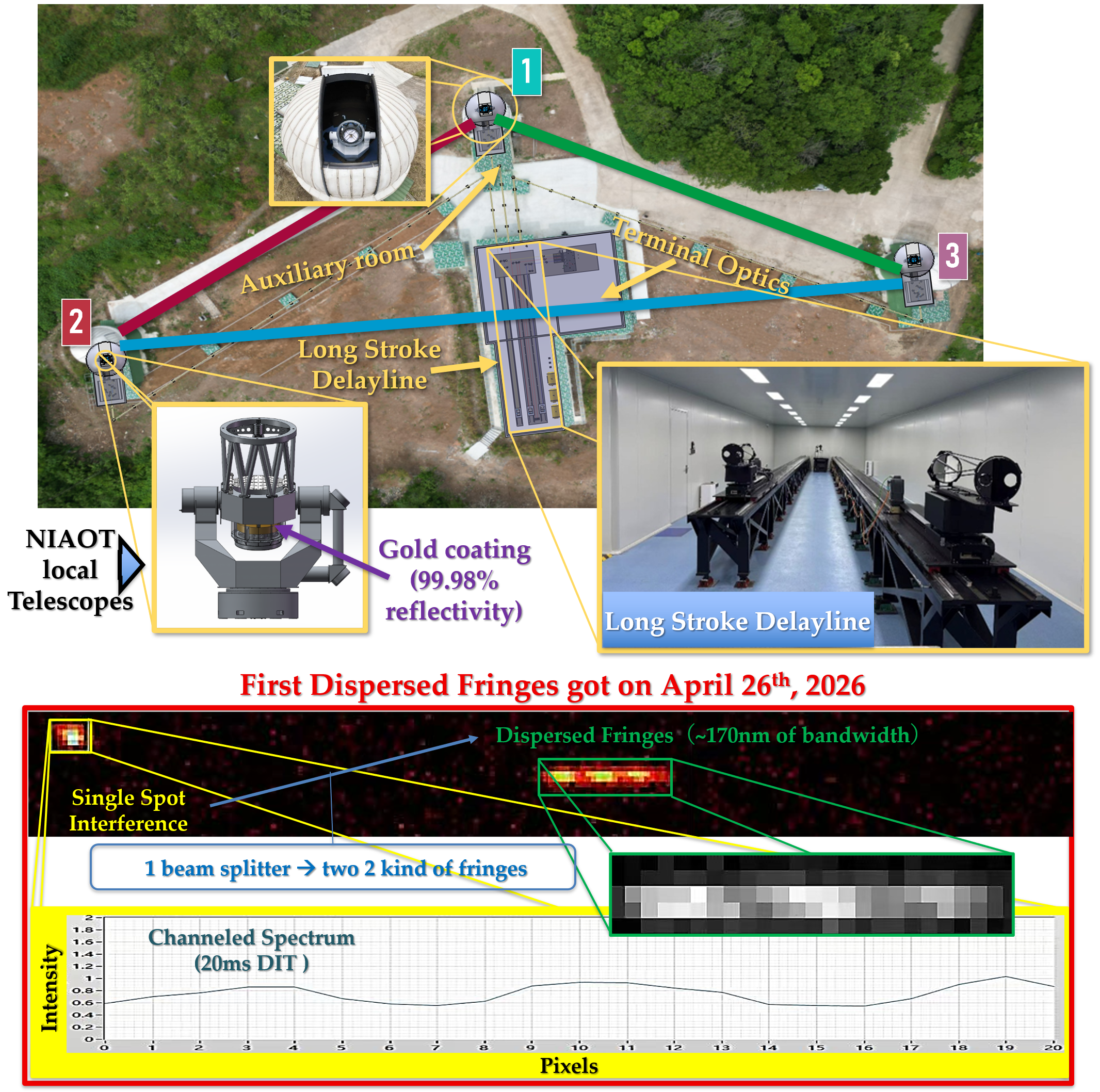}
\end{tabular}
\end{center}
\caption[example] { \label{fig:1} 
\textbf{\textit{Top:}} Overview of the XuYi OLBIN. \textbf{\textit{Bottom:}} First dispersed fringes got on April 26th, 2026.}
\end{figure}

Figure \ref{fig:1} provides a general overview of the XuYi Optical Long Baseline Interferometer (OLBIN), highlighting the NIAOT gold-coated telescopes (99.98\% reflectivity), their enclosures, the auxiliary buildings hosting the adaptive optics systems, the underground vacuum pipelines, and the central optical facility containing the long-stroke delay line and the beam-combination unit. The bottom panel displays the first dispersed interference fringes recorded on April 26, 2026. These data were obtained by splitting a channelled spectrum using a beam splitter. One arm directly produced a white-light interference spot (left side of the CCD), while the second arm was directed through a prism to generate spectrally dispersed fringes (right side of the CCD), demonstrating the successful detection and spectral dispersion of the interference signal.

\section{SOFTWARE ARCHITECTURE}
\label{sec:3}

The simulator was developed in MATLAB and incorporates a graphical user interface (GUI) designed to facilitate rapid observation planning and feasibility assessment.

Its architecture follows a modular design composed of four main components:

\begin{enumerate}
\item Target observability module,
\item Interferometric geometry module,
\item $(u,v)$-coverage module,
\item Visibility prediction module.
\end{enumerate}

The user can define the following observational and instrumental parameters:

\begin{itemize}
\item Observation date,
\item Target name and/or celestial coordinates,
\item Telescope opening angle,
\item Universal Time (UT) or local time,
\item Delay-line stroke length (up to 44,m),
\item Target angular diameter,
\item Observing spectral band,
\item Spectral resolution.
\end{itemize}

In addition, the software includes two temporary input parameters associated with the initial commissioning configuration:

\begin{itemize}
\item Side length of the square entrance aperture (m),
\item Height of the reflecting mirror above ground level (m).
\end{itemize}
These additional parameters are specific to the temporary siderostat configuration employed during the commissioning phase of the XuYi OLBIN. Before the installation of the permanent telescopes, the incoming stellar light was collected through a square aperture and redirected by a flat mirror toward the interferometric beam transport system. The simulator therefore allows the user to account for the geometrical constraints imposed by this provisional optical arrangement.

The software can operate using either local or universal time and automatically converts all coordinates into the XuYi site reference frame.

\begin{figure} [ht]
\begin{center}
\begin{tabular}{c} 
\includegraphics[height=10cm]{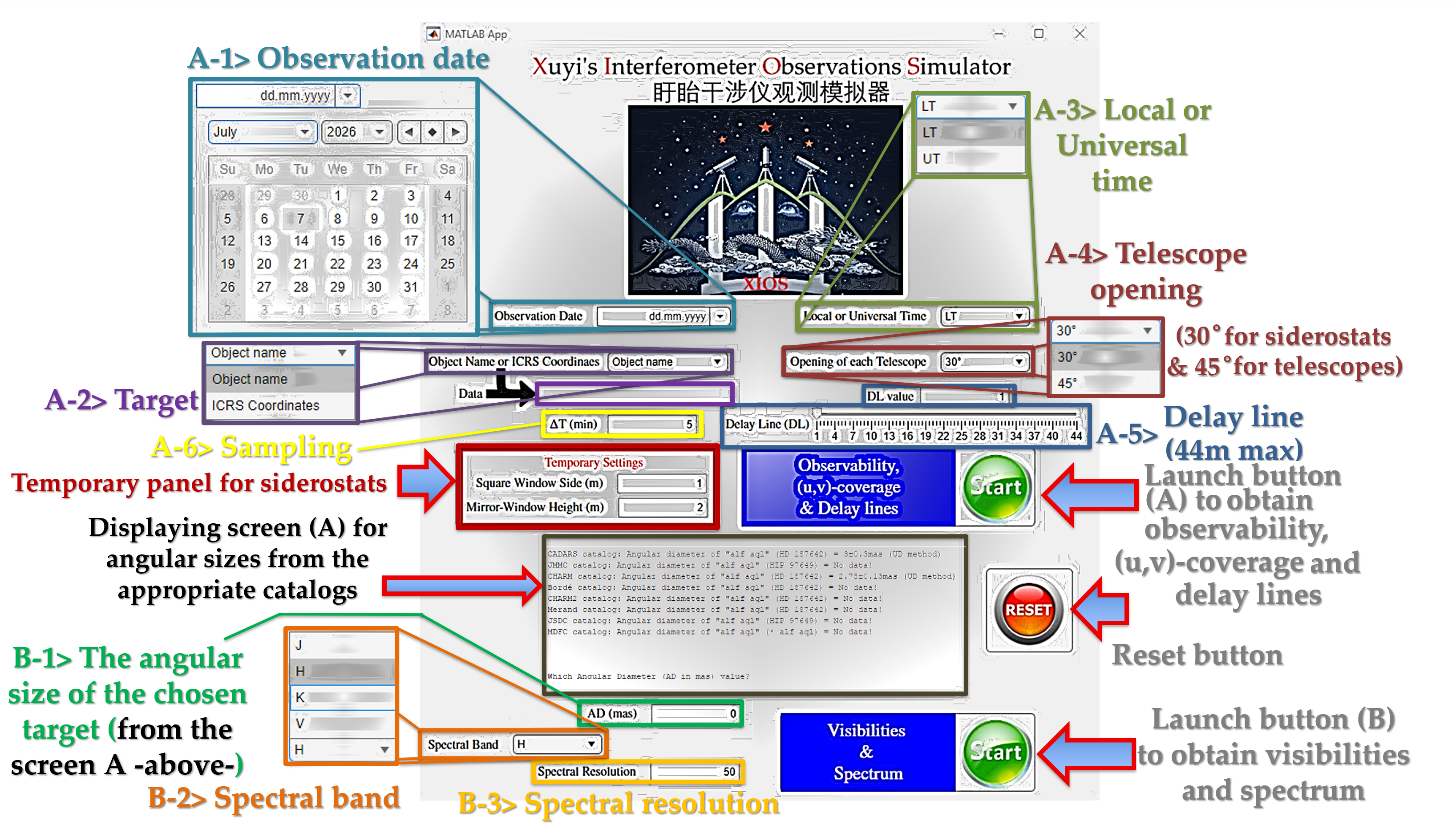}
\end{tabular}
\end{center}
\caption[example] { \label{fig:2} 
Graphical user interface (GUI) command panel of the XuYi Interferometric Observatory Simulator (XIOS).}
\end{figure}

Figure \ref{fig:2} shows the MATLAB GUI command panel of the XuYi Interferometric Observation Simulator (XIOS).

The GUI was intentionally designed to remain lightweight and fast, enabling rapid exploration of observing scenarios without requiring external databases or complex installation procedures.

\section{OBSERVABILITY COMPUTATION}
\label{sec:4}
\subsection{Daily observability}
\label{sec:4.1}

The daily observability calculation is based on routines from the MATLAB Astronomy and Astrophysics Toolbox (MAAT) \cite{ofek2014maat}.

For a selected target, the software computes:
\begin{itemize}
\item Altitude,
\item Azimuth,
\item Zenith angle,
\item Airmass,
\item Transit time,
\end{itemize}
throughout the night.

The resulting observability window is displayed similarly to the widely used ASPRO2 software \cite{2013ascl.soft10005B}.

\begin{figure} [ht]
\begin{center}
\begin{tabular}{c} 
\includegraphics[height=10cm]{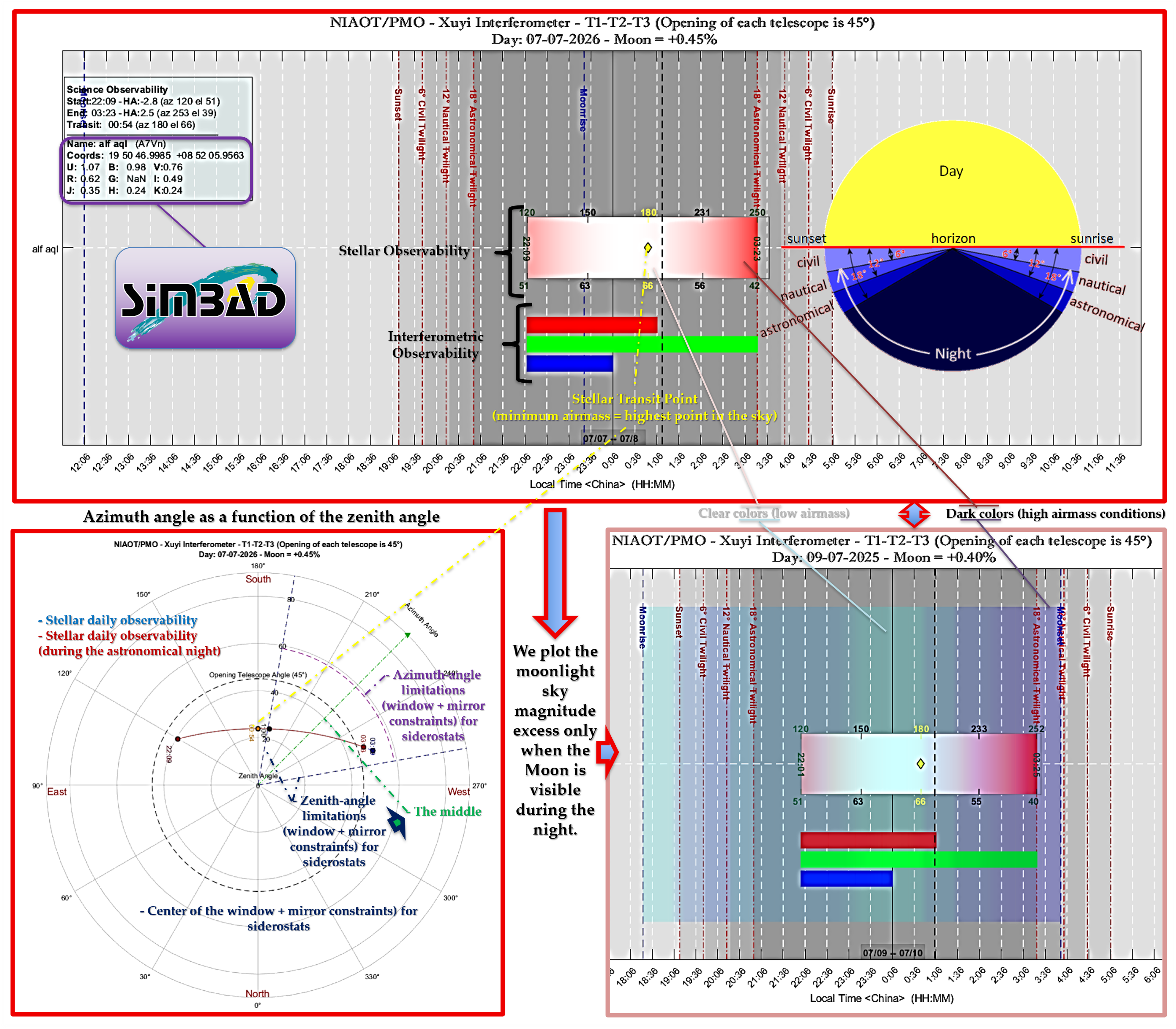}
\end{tabular}
\end{center}
\caption[example] { \label{fig:3} 
XIOS observability diagram of Altair showing twilight limits, Moon visibility, airmass evolution, and interferometric observing windows.}
\end{figure}

Figure \ref{fig:3} illustrates, as an example, the observability of Altair ($\alpha$ Aquilae) on July 7, 2026. The plot displays the target's observability as a function of time, together with the azimuth angle versus zenith angle. The moonlight-induced sky brightness excess is shown only when the Moon is above the horizon. For comparison, the same target is also evaluated one year earlier, on July 7, 2025. The title of this figure automatically provides the interferometer name and configuration, the telescope opening angle, the observation date, as well as the lunar illumination fraction and phase (increasing or decreasing), indicated by the sign preceding the percentage (e.g., +0.45 = 45\% illuminated, waxing; -0.45 = 45\% illuminated, waning).

All fundamental stellar parameters are automatically retrieved from the SIMBAD database and subsequently stored in a local database for future use.

The horizontal red dashed lines indicate sunset, sunrise, and the limits of civil (6\textdegree), nautical (12\textdegree), and astronomical (18\textdegree) twilight. The horizontal blue dashed lines correspond to Moonrise and Moonset. The colored rectangle represents the excess sky brightness caused by moonlight.

The main vertical rectangle indicates the target observability window, while the associated color gradient represents the evolution of the airmass throughout the night. The three additional vertical rectangles shown below correspond to the interferometric observability windows, which are further constrained by the delay-line limits of the interferometer (see Sec.~\ref{sec:4.2}). The color coding provides a direct visualization of atmospheric observing conditions through the temporal variation of airmass.

\subsection{Delay-line constraints}
\label{sec:4.2}

An interferometer cannot observe a target unless the Optical Path Difference (OPD) remains within the available delay-line stroke.

For each baseline, the simulator,through (u,v)-coverage equations (Sec.~\ref{sec:5}), computes the projected OPD:
\begin{equation}
\label{eq:2}
\mathcal{OPD} = \overrightarrow{B}\cdot\overrightarrow{s},
\end{equation}
where $\overrightarrow{B}$ is the baseline vector and $\overrightarrow{s}$ is the source direction vector. 

The resulting delay-line requirements are compared against the 44 m mechanical limitation currently implemented at XuYi.

Only intervals satisfying all baseline constraints are considered interferometrically observable, as shown in Fig.~\ref{fig:3} by the three horizontal bars (red: T1–T2, green: T1–T3, blue: T2–T3).

\section{(u,v)-COVERAGE SIMULATION}
\label{sec:5}

Earth-rotation synthesis, also called super-synthesis, was first introduced in the framework of radio interferometry by Martin Ryle and further developed in early pulsar studies with Antony Hewish (Nobel Prize in Physics, 1974) \cite{ryle1970,hewish1968}. It causes the projected baselines to evolve continuously during observations.

The simulator computes the corresponding spatial-frequency coordinates in Fourier space, through the topocentric interferometric coordinate transformation matrix $R$:
\begin{equation}
\label{eq:3}
(u,v,w) = R(H,\phi)\, \mathbf{B}, \quad \text{with projection onto } \mathbf{s}(\alpha,\delta),
\end{equation}
where $H$ is the hour angle, which depends on the local sidereal time and thus on the observer’s longitude $\lambda$, while right ascension $\alpha$ and declination $\delta$ define the fixed equatorial coordinates of the target. The transformation matrix $R(H,\phi)$ projects the baseline from the terrestrial reference frame into the $(u,v,w)$ interferometric frame, and depends on the observer’s latitude $\phi$ through the geometric projection.

The full derivation of the interferometric spatial-frequency formalism is given in Chapter 4 of *Interferometry and Synthesis in Radio Astronomy* by A. R. Thompson et al. (2001)\cite{thompson2001interferometry}, and the time interval is directly and automatically inferred from the target daily observability (Sec.~\ref{sec:4.1}).

The resulting trajectories are displayed both in:
\begin{itemize}
\item two-dimensional (u,v)-coverage, 
\item three-dimensional baseline representations, also called the (u,v,w)-coverage. 
\end{itemize}

The component $w$ corresponds to the delay term (Sec.~\ref{sec:4.2}). The software currently supports all three XuYi baselines simultaneously, allowing users to evaluate Fourier-plane sampling quality prior to observations.

Examples generated for bright stellar targets demonstrate the expected elliptical tracks characteristic of Earth-rotation synthesis.

Such information is critical for future image reconstruction studies.

\begin{figure} [ht]
\begin{center}
\begin{tabular}{c} 
\includegraphics[height=10cm]{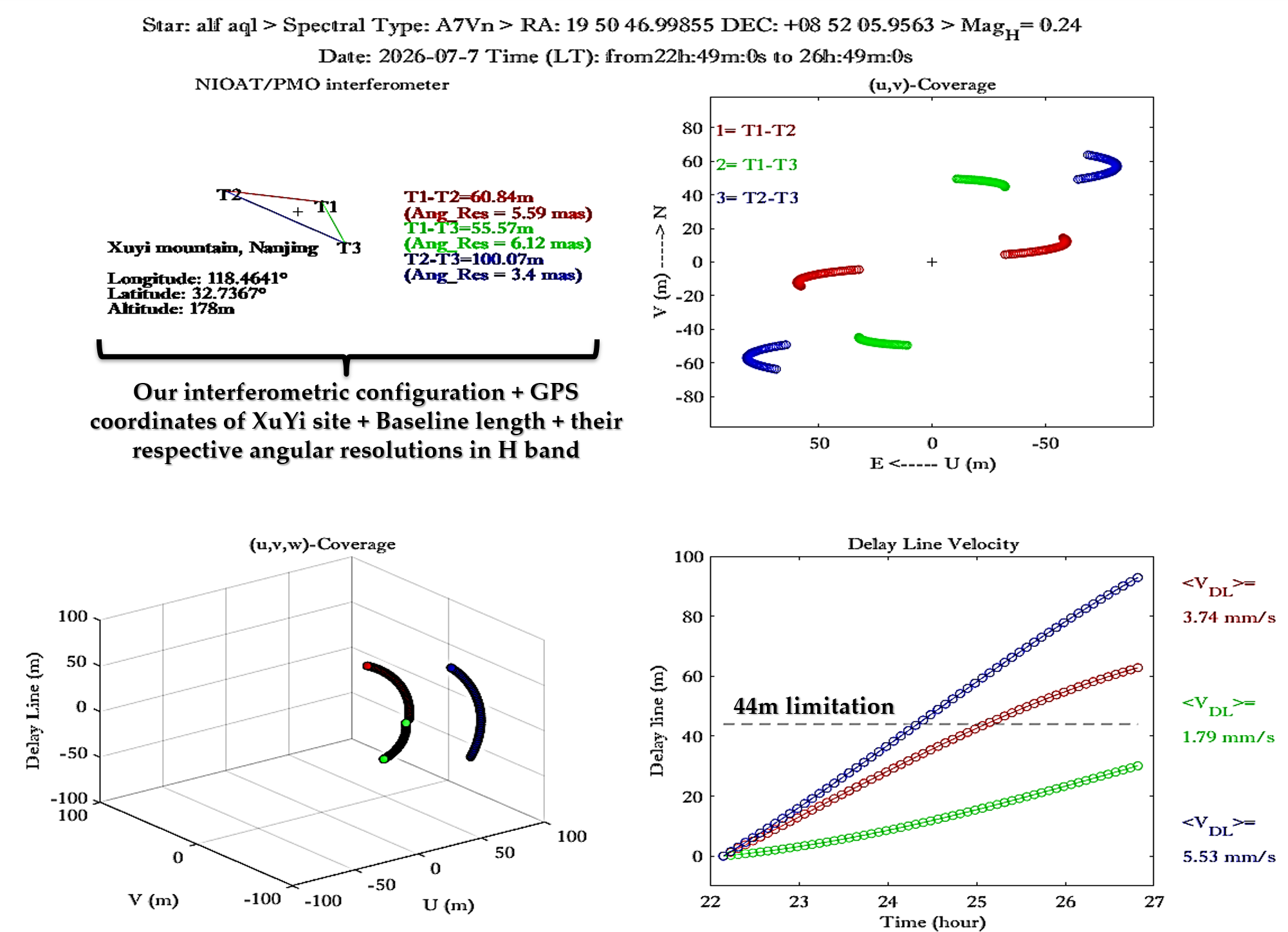}
\end{tabular}
\end{center}
\caption[example] { \label{fig:4} 
\textbf{\textit{Top-left:}} Interferometric configuration of the XuYi interferometer. \textbf{\textit{Top-right:}} $(u,v)$-coverage. \textbf{\textit{Bottom-left:}} $(u,v,w)$-coverage. \textbf{\textit{Bottom-right:}} Delay-line behaviour and corresponding average velocities. The dashed horizontal line indicates the maximum delay-line stroke (44 m) and its effect on the interferometric baselines, as shown in the previous Fig.~\ref{fig:3}.}
\end{figure}

Figure~\ref{fig:4} shows, for Altair on July 7, 2026, the interferometric configuration of the XuYi interferometer, the $(u,v)$ and $(u,v,w)$ coverages, as well as the delay-line behaviour. The title of this figure automatically provides the target name, its spectral type, right ascension, declination, H-band magnitude, observation date, as well as the time interval of target observability for that night.

\section{VISIBILITY ESTIMATION}
\label{sec:6}
\subsection{Angular diameter}
\label{sec:6.1}

In order to compute the visibility, the angular diameter of the target must first be determined. For this purpose, we use several stellar angular diameter catalogues. The angular diameters can be obtained from seven predetermined catalogues, namely:
\begin{enumerate}
\item CADARS (Catalogue of Apparent Diameters and Absolute Radii of Stars) \cite{cadars2001},
\item CHARM (Catalogue of High Angular Resolution Measurements) \cite{charm2002},
\item Bordé et al. (calibrator star catalogue, named after the first author) \cite{borde2002},
\item CHARM2 (Catalogue of High Angular Resolution Measurements, version 2) \cite{charm2_2005},
\item Mérand et al. (calibrator star catalogue, named after the first author) \cite{merand2005},
\item JSDC (JMMC Stellar Diameters Catalogue) \cite{bonneau2006jsdc},
\item MDFC (Mid-infrared Stellar Diameters and Fluxes Compilation Catalogue) \cite{cruzalebes2024mdfc}.
\end{enumerate}

The angular diameter of the target, in milliarcsecond (mas), must be entered manually after being selected from the available catalogue results displayed on the main GUI screen.

Note that the different angular diameter catalogues provide simple geometric source models, including:
\begin{itemize}
\item Uniform Disk (UD),
\item Limb-darkened Disk (LD).
\end{itemize}

\subsection{Visibility calculations}
\label{sec:6.2}

From the angular diameter, the visibility is computed as:
\begin{equation}
\label{eq:4}
V(B,\lambda) = \frac{2J_{1}(x)}{x}, \quad \text{with} \quad x = \frac{\pi \theta B}{\lambda}.
\end{equation}

where $J_{1}$ is the first-order Bessel function, $\theta$ is the angular diameter, $B$ is the projected baseline, and $\lambda$ is the observing wavelength.

The simulator produces spectral visibility distributions for each projected baseline as a function of wavelength. The chosen spectral resolution determines the number of spectral visibility points.

These outputs allow observers to assess whether a target is resolved, partially resolved, or unresolved prior to scheduling observations.

\begin{figure} [ht]
\begin{center}
\begin{tabular}{c} 
\includegraphics[height=7cm]{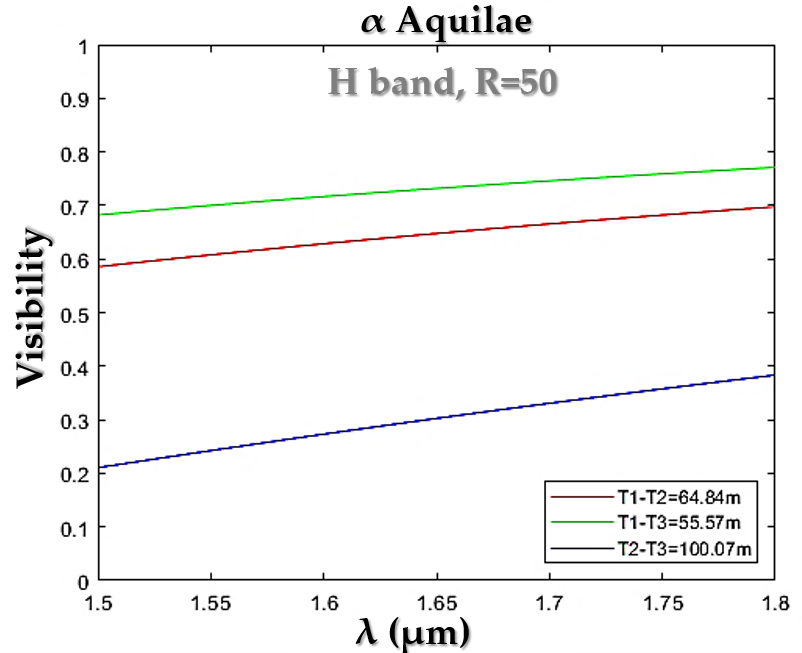}
\end{tabular}
\end{center}
\caption[example] { \label{fig:5} 
Visibilities versus wavelength for Altair on July 7, 2026, for the triplet baselines T1T2 (red), T1T3 (green), and T2T3 (blue), in the H band with a spectral resolution of $R=50$.}
\end{figure}

Figure~\ref{fig:5} depicts, for Altair on July 7, 2026, the visibilities versus wavelength for the three baselines T1T2, T1T3, and T2T3 in the H band, with a spectral resolution of $R=50$.

\section{Spectrum}
\label{sec:7}

\begin{figure}[ht]
\centering
\includegraphics[height=3cm]{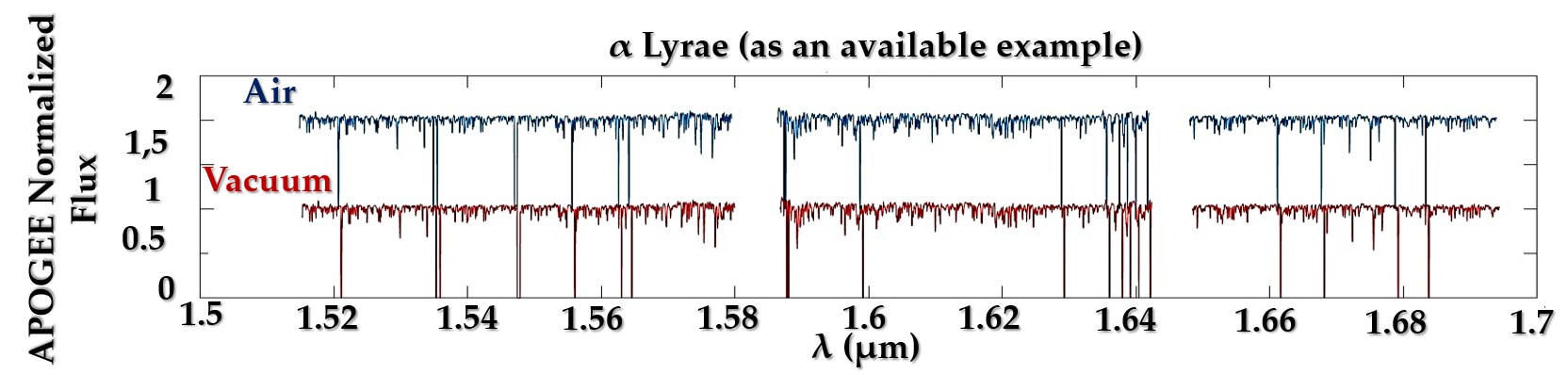}
\caption[example]{\label{fig:6}
SDSS/APOGEE spectrum of Vega in the H band, in air (blue) and vacuum (red).}
\end{figure}

XIOS also displays the spectrum of the selected target, as shown in Fig.~\ref{fig:6}. We use SDSS/APOGEE data, which are very limited in the H band. Since Altair is not available in these databases, we instead show Vega ($\alpha$ Lyrae) as an example.

Due to the lack of ground-based spectral databases in the H band, observational spectra should be complemented by space-based spectral databases and/or synthetic spectra derived from stellar atmosphere models (e.g. Tlusty/SYNSPEC, MARCS, etc.).

Several complementary uses of XIOS are currently available, namely:
\begin{itemize}
\item OIFITS export capability for other stellar modeling applications and image reconstruction,
\item multiple beam-combiner configurations,
\item fast rotator, hot active stars (e.g. with SCIROCCO \cite{2014A&A...569A..45H,2018MNRAS.480.1263H,2022MNRAS.511.4724H}), cold evolved stars (e.g. with PAMPERO \cite{2019MNRAS.489.2595H}), binary-star, and circumstellar disk models.
\end{itemize}

Several upgrades are also currently planned:
\begin{itemize}
\item atmospheric turbulence simulation,
\item fringe-tracking performance estimation,
\item direct interoperability with ASPRO2 and SearchCal,
\item integration of future XuYi telescope stations.
\end{itemize}

Additional functionalities dedicated to image reconstruction preparation are also under consideration.

\section{CONCLUSIONS}
\label{sec:conclu}

We have developed the XuYi Interferometer Observation Simulator, a MATLAB-based software package dedicated to the preparation of observations for the first Chinese optical long-baseline interferometer.
The simulator combines target observability, delay-line constraints, Earth-rotation synthesis, (u,v)-coverage generation, and visibility estimation within a unified graphical environment.
Its lightweight architecture allows rapid observation planning while providing physically realistic predictions of interferometric performance.
The tool constitutes an important step toward future scientific exploitation of the XuYi Interferometer and will continue to evolve alongside the development of the facility.

\bibliography{report} 
\bibliographystyle{spiebib} 

\end{document}